\documentclass[preprint,showpacs,preprintnumbers,amsmath,amssymb]{revtex4}

\usepackage{graphicx}
\usepackage{dcolumn}
\usepackage{bm}

\begin{document}

\preprint{Submitted to Physical Review E, 2008}

\title{A Markov Process Inspired Cellular Automata Model of Departure Headways}

\author{Fa Wang, Li Li, Jianming Hu, Yan Ji, Danya Yao, Yi Zhang, Xuexiang Jin, Yuelong Su, Zheng
Wei}
 \altaffiliation{Department of Automation, Tsinghua University, Beijing China
 100084}
 \email{li-li@mail.tsinghua.edu.cn}

\date{received: March 18, 2008; revised: \today}

\begin{abstract}
To provide a more accurate description of the driving behaviors in
vehicle queues, a namely Markov-Gap cellular automata model is
proposed in this paper. It views the variation of the gap between
two consequent vehicles as a Markov process whose stationary
distribution corresponds to the observed distribution of practical
gaps. The multiformity of this Markov process provides the model
enough flexibility to describe various driving behaviors. Two
examples are given to show how to specialize it for different
scenarios: usually mentioned flows on freeways and start-up flows at
signalized intersections. The agreement between the empirical
observations and the simulation results suggests the soundness of
this new approach.
\end{abstract}


\maketitle

\section{Introduction}
\label{intro}

Traffic flow is a many-body system of strongly interacting vehicles.
Various models are presented to understand the observed rich variety
of physical phenomena exhibited by traffic flow
\cite{ChowdhurySantenSchadschneider2000}, \cite{Helbing2001},
\cite{MahnkeKaupuzsLubashevsky2005}. From the viewpoint of vehicles'
queuing dynamics, these models generally depict four kinds of
traffic flows: 1) static queues (e.g. vehicles parked on a line
\cite{Lee2004}, \cite{RawalRodgers2005},
\cite{KernerKlenovHillerRehborn2006}, \cite{Abul-Magd2006}, and
vehicle queues fully-stopped in front of signalized intersections
\cite{JinSuZhangLi2008}), 2) stable moving queues (vehicle platoons)
which includes the well-known Kerner's synchronized flow and
wide-moving jams \cite{Kerner2004}, 3) free-flows formulating no
explicit queues, and 4) unstable queues which contains complex
inter-arrival and inter-departure queuing interactions
\cite{HelbingTreiberKesting2006}, \cite{SchonhofHelbing2007}.

The queues with complex interactions between its elements attract
increasing attentions recently. In many known approaches, the
dynamics of flows are described on $N$ strongly-linked particles
under fluctuations \cite{Helbing2001},
\cite{MahnkeKaupuzsLubashevsky2005}. Since the governing interaction
forces or potentials are not directly measurable for traffic flow
applications, the statistical distributions of particles are often
investigated instead. For example, the important phase transition
phenomena of traffic flow were studied by using scattering theory in
\cite{HelbingTreiberKesting2006}, \cite{KrbalekHelbing2004}.
Differently in \cite{Abul-Magd2007}, random matrix theory was
applied to predict the space-gap distribution between vehicles in
three-phrase flows. However, these studies focus on the steady-state
statistics. Two questions thus arise as: 1) can we design
microscopic simulation models (e.g. car-following, cellular
automata) that meanwhile yields such statistical properties; 2) how
to depict the transient-state statistics of inter-arrival and
inter-departure queuing interactions. This paper aims to propose a
simple yet useful CA model to partly answer these two questions.

In the field of traffic flow modeling, cellular automata (CA) based
microscopic traffic simulation has received constant interests in
the last decade, because of its efficiency on fast simulation and
ability to imitate the driving behaviors to explain some complex
phenomena \cite{MaerivoetMoor2005}. However, many CA models are
designed to study the phase transitions of traffic flow only, i.e.
\cite{NagelScheckenberg1992}, \cite{HelbingSchreckenberg1999}. The
rather low accuracy on vehicles' velocity and acceleration, which is
useful in computer implementations, prevents them to incorporate
further more features.

To deal with this problem, a new CA model is proposed in this paper.
It models the variation of the gap between two consequent vehicles
as Markov process, whose stationary distribution can be designed to
reflect the distribution of empirical gap data. Because it is a
dynamic model, it can simulate the inter-arrival and inter-departure
queuing interactions, too. Moreover, the velocity adjusting rule can
be gotten in a concise and unified way, which contains no
intractable formulations and fits for fast simulations.

To prove its soundness, this general model is specialized to explain
the interesting phenomena of traffic breakdown of freeway traffic
flow and departure headways at signalized intersections. The
agreement between the empirical and simulation results indicates
that this new model approvingly reveals the implicit interactions
between queuing vehicles. The results also prove the statement given
in \cite{HelbingTreiberKesting2006} that the interdeparture time
statistics is dominated by spatial interactions among the elements
in a queue rather than the arrival or exit processes for a
relatively long queue.

\section{The Markov-Gap CA Model}
\label{sec:1}

\begin{figure}[h]
\resizebox{0.75\columnwidth}{!}{%
  \includegraphics{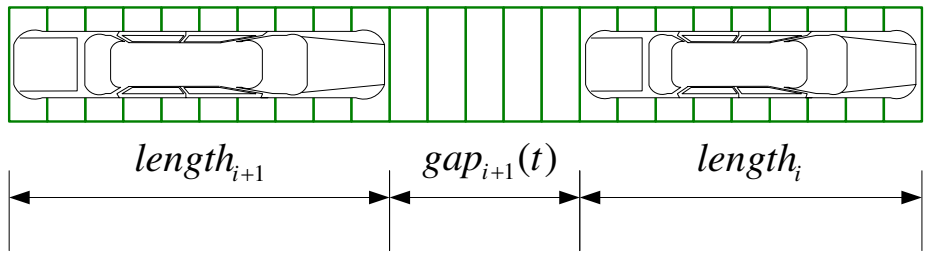}
}
\caption{Diagram of the mutil-cell CA model.}
\label{fig:1}       
\end{figure}

The model proposed here is a multi-cell CA model on a
unidirectional, single-lane lattice, in which a vehicle is allowed
to span a number of consecutive cells in the longitudinal direction;
see Fig.~\ref{fig:1}. Multi-cell type is chosen here for modeling
the possible small gaps during slow moving scenarios.

%

Usually, CA models of road traffic contain two modes: free-driving
mode, following/ braking mode. In the following/braking mode, a
driver adjusts his/her vehicle to keep an as close as possible but
not too close distance gap. Due to various reasons (inability to
predict the leader's action or make precise driving actions,
possibility to be absent minded, etc.), he/she cannot maintain a
constant gap. Suppose altogether we have $N$ possible gaps whose
lengthes range from $0$ cell to $(N-1)$ cells. If at time $t$, the
length of the gap is $n$; at the next time $(t+1)$, the length may
still remain $n$, or possibly change to an adjacent value, i.e.
$(n-1)$, $(n+1)$, etc.

Abstractly, this variation process of a gap during following can be
viewed as a transferring process between $N$ states, where each
state denotes a distinct gap in length. Statistically, considering
the relatively long-time driving behaviors of many drivers, the
transferring probability from state $i$ to $j$ would approach a
steady value. That is, the drivers' car-following actions can be
depicted by a Markov Chain characterizing the variation process of
gap. To our best knowledge, only \cite{SonKimLee2005} proposed a
state-transition model similar to this model in a vague way, but it
does not mention the Markov transferring probability.

Notice that the above Markov Chain is positive recurrent and
aperiodic, it yields a steady-state (stationary) distribution, which
straightforwardly corresponds to the empirical distributions of gap
distance in traffic flow applications \cite{Luttinen1992},
\cite{MichaelLeemingDwyer2000}. This makes the model able to yield
expected statistical features on gap/headways.

Indeed, it is not required to discretize the steady-state
distribution into the minimal division level (the length of a cell
here), because to consider too many transition states would be
unnecessary and time consuming. In this paper, the so called Markov
Chain Aggregation technique used in \cite{Stewart1994},
\cite{Spears1999}, \cite{BolchGreinerdeMeerTrivedideMeerTrivedi2006}
is employed to combine some neighboring states and thus simplify
transition matrix. Without losing much generality, we can further
assume that a state (may span a number of consecutive cells) can
only transfer to itself or the strictly neighboring state (noticing
the simulation time span is usually short and the speed is limited,
this assumption is reasonable). Thus, the transition matrix of the
Markov Chain can be written into a tridiagonal matrix as:

\begin{equation}
\label{equ:1}\left[\begin{array}{cccccc}

r_0 & q_0 & 0 & 0 & \cdots & 0\\
p_1 & r_1 & q_1 & 0 & \cdots & 0\\
0 & p_2 & r_2 & q_2 &  & \vdots \\
\vdots & & &  &  & \vdots \\
0 & \cdots & \cdots & 0 & p_{n-1} & r_{n-1} \\

\end{array}\right]
\end{equation}

\noindent where $p_i$, $r_i$, $q_i$ are the probabilities that
transfer from state $i$ to the lower, current and higher states,
respectively.

Let $p_0 = q_{n-1} = 0$, we have

\begin{equation}
\label{equ:2} p_i + r_i + q_i = 1
\end{equation}

\noindent for $i \in \{0, 1, ..., N-1 \}$.

As well known, there exist numerous solutions for the potential
transition matrix (\ref{equ:1}). Clearly any a valid solution could
be used for simulation, but we generally have two more requirements
for constructing Markov Chain for simulations: 1) lower time
complexity of the formulation algorithm; and 2) faster convergency
speed of the obtained Markov Chain, since it determines how many
rounds of simulations are needed before drawing a satisfactory
conclusion \cite{Stewart1994}, \cite{MeynTweedie1994}. To meet these
two requirements, we can formulate such a matrix construction
problem into a Semidefinite Programming (SDP) problem (e.g. SDP-(19)
in \cite{BoydDiaconisXiao2004}), which is easy to solve. In the rest
of this paper, let's assume the associated transition matrices
(\ref{equ:1}) have already been obtained, since how to derive such a
matrix does not accord with the main theme here.

Suppose the gap between the $(i-1)$th and $i$th vehicle is in the
$n$th state at time $t$, which indicates that the gap of the $i$th
vehicle satisfies $g_{i}(t) \in [g_{n}^{low}, g_{n}^{up}]$. Here,
$g_{n}^{low}$ and $g_{n}^{up}$ denote the lower and upper boundaries
of the $n$th state, respectively.

Based on the above Markov transition idea, the gap might be enlarged
into the $(n+1)$th state with probability $q_{n}$ due to
deceleration. In the simulation, an arbitrary random number
$\widetilde{g_i^+}(t+1)$ will be uniformly generated in
$[g_{n+1}^{low}, g_{n+1}^{up}]$ and assigned as the potential
forthcoming gap. And the expected velocity $v_i$ of the $i$th
vehicle at time $(t+T)$ should be

\begin{equation}
\label{equ:3} \widetilde{v_i^-}(t+T) = v_{i-1}(t) -
[\widetilde{g_i^+}(t) - g_i(t)] / T
\end{equation}

Similarly, when acceleration, the velocity should be adjusted as

\begin{equation}
\label{equ:4} \widetilde{v_i^+}(t+T) = v_{i-1}(t) -
[\widetilde{g_i^-}(t) - g_i(t)] / T
\end{equation}

\noindent where $\widetilde{g_i^-}(t) \in [g_{n-1}^{low},
g_{n-1}^{up}]$.

Noticing most CA models of road traffic consists of an accelerating
rule to make the vehicle approach the highest possible speed and a
braking rule to avoid collisions, we can similarly write the whole
updating rules as:

1) If $g_i(t) > G_{max}$, where $G_{max}$ denotes the maximum
coupling distance; the vehicle is in free-driving scheme, then it
will approach the highest velocity as

\begin{equation}
\label{equ:5} v_{i}(t+T) = \min \{v_{i}(t) + a_{max}^+ \times T,
v_{max} \}
\end{equation}

Else if $v_i(t) - v_{i-1}(t) > [g_i(t) - G] / H$, there exists a
risk to collide, and thus let the vehicle brake

\begin{equation}
\label{equ:6} v_{i}(t+T) = v_{i}(t) - D \times T
\end{equation}

\noindent where $G$, $H$ and $D$ are constants denoting the minimum
safety gap, deceleration time and braking decelerating rate,
respectively.

Otherwise, the velocity of the vehicle is updated as follows, where
the $\max$, $\min$ functions are added to guarantee that the
velocity and ac/decelerating rates are within the limits.

\begin{equation}
\label{equ:7} v_i(t + T) = \left\{ \begin{array}{ll}

\min \{v_i(t) + a_{max}^+ \times T, \widetilde{v_i^+}(t+T), v_{max} \} & \textrm{, with $p_{n}$} \\
v_i(t) & \textrm{, with $r_{n}$} \\
\max \{v_i(t) - a_{max}^- \times T, \widetilde{v_i^-}(t+T), 0 \} &
\textrm{, with $q_{n}$}

\end{array} \right.
\end{equation}

2) The position of vehicle is then updated as

\begin{equation}
\label{equ:8} x_i(t + T) = x_i(t) + v_i(t) \times T
\end{equation}

Simulations show that the multiformity of this Markov process
provides the model enough flexibility to describe diversified
driving behaviors. The transferring matrix can be explicitly
determined by the investigation data from drivers or implicitly
estimated through the observed distribution of gap.

However, the above model should not be directly applied, if the
velocity of vehicles varies in a relatively large range. This is
primely because drivers would like to keep a roughly constant time
headway instead of gap distance. (As shown in
\cite{KernerKlenovHillerRehborn2006}, \cite{Abul-Magd2007}, the
time-headway distributions computed for the free-flow, synchronized
flow, and moving jam traffic phases do not differ too much in shape
and mean values.) Usually, the larger the velocity is, the larger
the gap distance should be. Thus, the aforementioned transition
matrix (\ref{equ:1})) should be modified to be velocity-dependent in
such situations. Examples of implementing this trick will be shown
in the coming sections.

\section{A Specialization for Flows on Freeways}
\label{sec:3}

To illustrate the effectiveness of the proposed model, a specialized
form is used to reproduce spontaneous traffic breakdowns and moving
jams that have been observed for flows on freeways. The velocity of
a vehicle may vary significantly in such situations. However, to
assign a Markov matrix (\ref{equ:1}) for each velocity (more
precisely to make the model into a time-dependent continuous time
Markov process) is troublesome for fast simulations; thus the
aggregation idea is used again.

\begin{figure}[h]
\resizebox{0.75\columnwidth}{!}{%
  \includegraphics{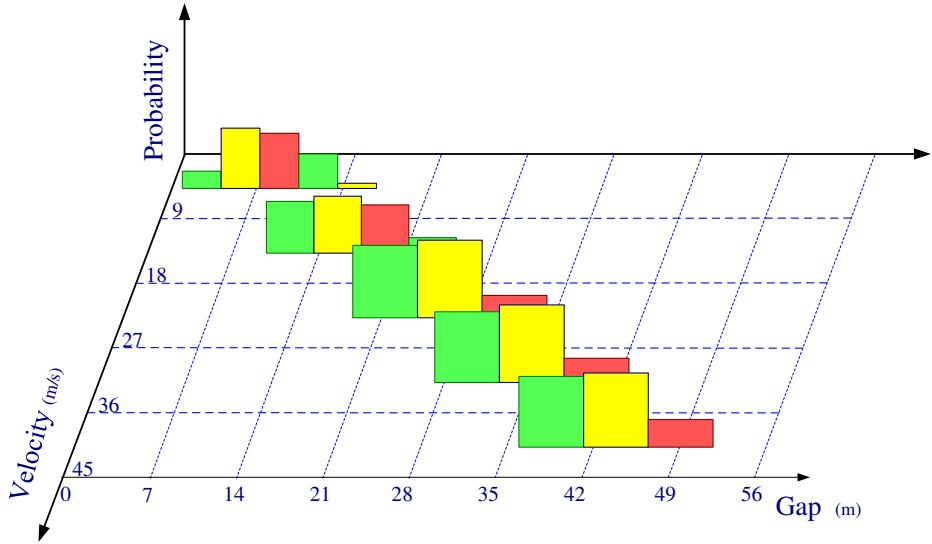}
}
\caption{The assumed discretized steady-state gap distributions
according to different velocities.}
\label{fig:2}       
\end{figure}

Suppose we only consider $J$ Markov Chains, each of which
corresponds to a certain range of velocity. When the velocity of the
$i$th vehicle $v_i(t)$ fall into the $j$th velocity range at time
$t$, we will apply the $j$th Markov transition matrix in velocity
updating. In the following simulation, five transition matrices are
designed (in a sense to the five speed stages of a car); see
Fig.~\ref{fig:2}. All the five distributions are uniformly
discretized histogram ($5$ segments, $3.3$m per segment for velocity
range $[0,9]$; $4$ segments, $4.1$m per segment for velocity range
$[9,18]$; and $3$ segment, $5.5$m per segment for the rests). The
discretized stationary distributions are assigned as
Table.~\ref{tab:table1} with respect to the practical observations
\cite{HelbingTreiberKesting2006}, \cite{SchonhofHelbing2007},
\cite{Abul-Magd2007}. The associated $v_{max}$ in (\ref{equ:5}) and
(\ref{equ:7}) are also set the $9$m/s, $18$m/s, ..., for each range
respectively.

%

\begin{table}[h]
\caption{\label{tab:table1}The discretized gap distributions
according to different velocities.}
\begin{ruledtabular}
\begin{tabular}{cccc}
Index & Velocity range(m/s) & Gap range(m) & Probability in each segment \\
\hline
1 & [0,9] & [0.5,17] & [0.10,0.35,0.32,0.20,0.03] \\
2 & [9,18] & [9,25.5] & [0.30,0.33,0.28,0.09] \\
3 & [18,27] & [18,34.5] & [0.42,0.45,0.13] \\
4 & [27,36] & [27,43.5] & [0.41,0.0.45,0.14] \\
5 & [36,54] & [36,52.5] & [0.41,0.43,0.16]
\end{tabular}
\end{ruledtabular}
\end{table}

Here, the length of each cell is set as $0.1$m. The length of each
vehicle is uniformly set as 4m. The maximum coupling distance
between two consecutive vehicles is set as $52.5$m, which means the
follower will enter free navigation mode if the gap to its leader is
greater than $52.5$m.

The velocity is updated once $T=1$s. The maximum velocity of a
vehicle is set to $v_{max}=45$m/s, the maximum accelerating rate is
$a_{max}^+=7.5$m/s$^2$, and the maximum decelerating rate is
$a_{max}^-=15$m/s$^2$. A fully stopped vehicle will start to move
only if the gap ahead is greater than $2.2$m. And the braking
parameters are set as $G=0.5$m, $H=7$s, $D=15$m/s$^2$.

%
\begin{figure}[h]
\resizebox{0.75\columnwidth}{!}{%
  \includegraphics{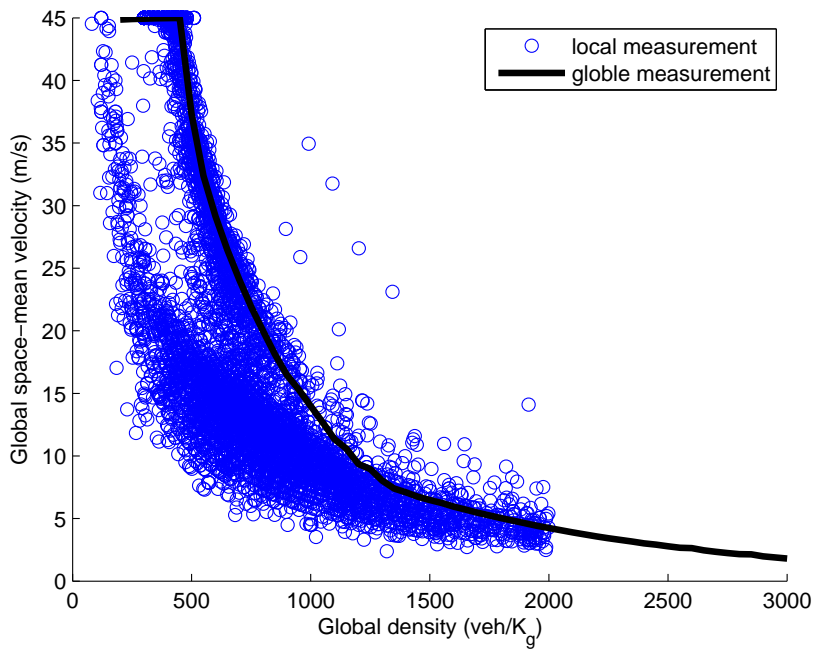}
}
\caption{The $(k, \bar{v}_s)$ diagram for the proposed model,
obtained by local and global measurements.}
\label{fig:3}       
\end{figure}

%
\begin{figure}[h]
\resizebox{0.75\columnwidth}{!}{%
  \includegraphics{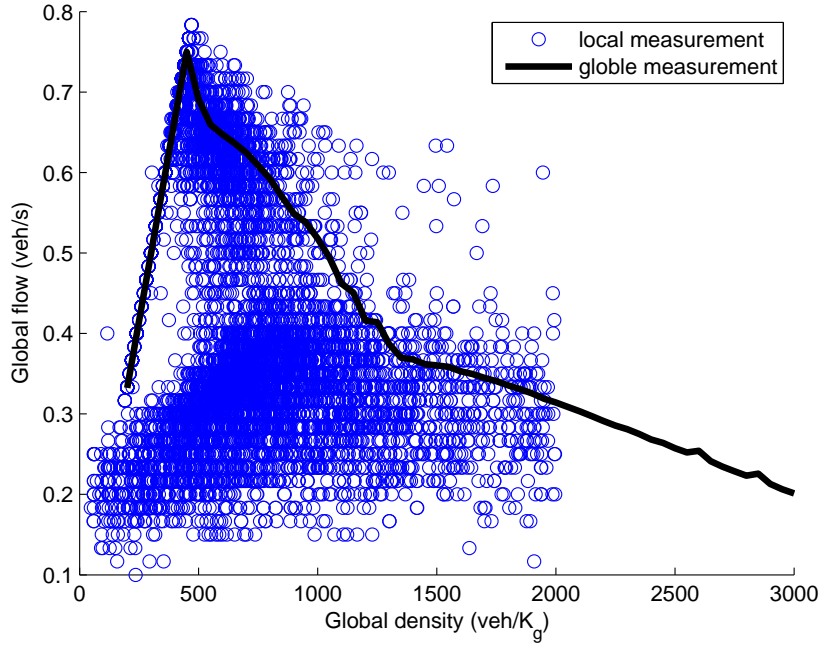}
}
\caption{The $(k, q)$ diagram for the proposed model, obtained by
local and global measurements.}
\label{fig:4}       
\end{figure}

Fig.~\ref{fig:3} and Fig.~\ref{fig:4} show the $(k, \bar{v}_s)$ and
$(k, q)$ diagrams for the proposed model, obtained by local and
global measurements. These microscopic traffic stream
characteristics are measured according to \cite{MaerivoetMoor2005}.
Here, $K_g$ denotes the length of the closed single-lane system,
which is $27000$m. It can be seen from Fig.~\ref{fig:3} that the
local measurements interestingly tend to form clusters around
certain space-mean velocities, which corresponds to the so called
optimal velocities predicated in \cite{Bando1995},
\cite{HelbingTilch1998}, \cite{JiangWuZhu2001}. These clusters
appear as branches in Fig.~\ref{fig:4}, too. This indicates that the
most important merit of the optimal velocity models has been
inherited by the proposed model, though via a different
interpretation of the dynamical equilibrium between traffic flow
velocity and density. Moreover, it yields the right headway
distribution as observed other than that given in Fig.7 of
\cite{Bando1995}.

Besides, the local measurements in Fig.~\ref{fig:4} also
discriminate Kerner's three-phase flows: the free-flow regime
contains only a few data points on line starts from the original
point; the synchronized regime is visible as a wide scatter of the
data points, having various speeds but relatively high flows; and
synchronized-flow, and jammed regimes contains the data points in
the wide-moving jam correspond to Kerner's line $J$. Thus, the
proposed model can reproduce the traffic phenomena of phase
transitions formation observed in real traffic.

It should be pointed out that the highest flow rate of the proposed
model is larger than that of NaSch model
\cite{NagelScheckenberg1992}, mainly because of the different
settings of vehicle's length.

\section{Another Specialization for Start-up Flow at Signalized Intersections}
\label{sec:4}

Another specialization is proposed in this section to explain the
distributions of the observed departure headway. Results indicate
that the proposed model can also depict the transient-state
statistics of unstable queues.

\begin{figure}[h]
\resizebox{0.75\columnwidth}{!}{%
  \includegraphics{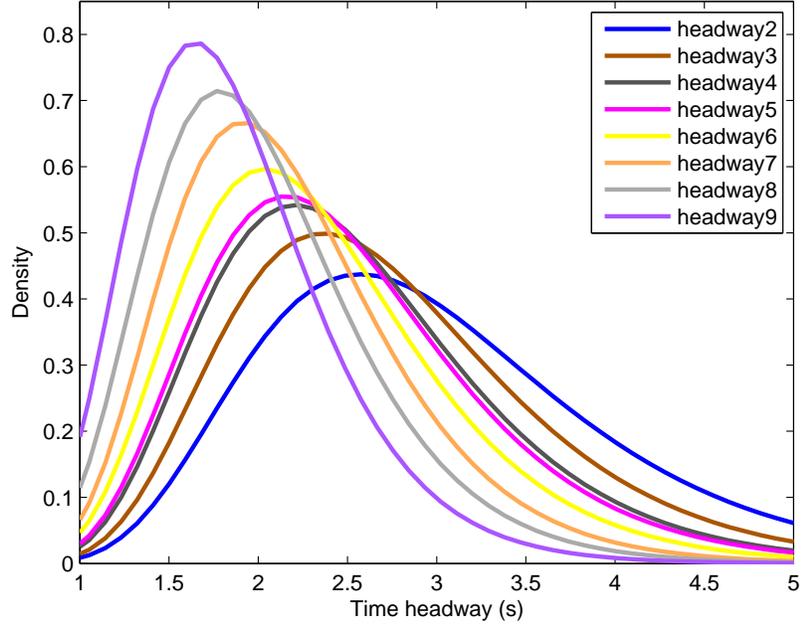}
}
\caption{The empirical distributions of departure headways
identified from the histograms observed. The departure headway data
used here are collected from several intersections in Beijing, China
during 2006 to 2007 by using video cameras. The recording time is
from 9:00 AM to 10:00 AM, when the traffic flow is not too crowded.
The queuing length does not grow so long (typically $5 \sim 9$
vehicles) in this period.}
\label{fig:5}       
\end{figure}

Departure headways are usually defined as the times that elapse
between consecutive vehicles when vehicles in a queue start crossing
the stop line (or any other reference line) at a signalized
intersection after the light turns green. Modeling departure
headways at signalized intersections attracts constant research
efforts due to its importance. Most previous works focus on the mean
departure headways of the first tens of vehicles. The results show
that the headways tend to decrease sequentially as queue position
increases, and a steady headway will be reached since the fourth or
the fifth vehicle, i.e. \cite{Luttinen1992},
\cite{MichaelLeemingDwyer2000}, \cite{ChangZhangMaoGong2005}.
Further investigations shows that each of these departure headways
follows a log-normal distribution (whose logarithm is normally
distributed) respectively and their mean values decrease to a
saturation value gradually \cite{LiWang2006},
\cite{SuWeiChengYaoZhangLiZhangLi2008}; see Fig.~\ref{fig:5}.

%

Only the movements of the first few vehicles, which are fully
stopped when the light turns green, are recorded and examined. The
vehicles whose length is too long (more than 10m) are discarded as
well as their following vehicles. The sample sizes decrease
according to the positions due to this data pre-processing process.
The statistics of the observed departure headways are listed in
Table.~\ref{tab:table2}. Kolmogorov-Smirnov (K-S) hypothesis test is
used here to check whether the empirical distributions are similar
to the estimated log-normal distributions, and Jarque-Bera (J-B)
hypothesis test is used to check the departure of the distribution
of empirical headways' logarithm from normality \cite{Gardiner1983}.
The criteria level is set as 95\%, which means the hypothesis will
be accepted if the value is greater than 0.05.

\begin{table}[h]
\caption{\label{tab:table2}The Kolmogorov-Smirnov (K-S) and
Jarque-Bera (J-B) hypothesis testing results of empirical departure
headways. $\mu$ and $\sigma$ denote the mean and standard deviation
of the logarithm of data respectively.}
\begin{ruledtabular}
\begin{tabular}{cccccc}
Postion & Sample size & K-S test & J-B test & $\mu$ & $\sigma$ \\
\hline
2 & 423 & 0.4282 & 0.2730 & 1.0569 & 0.3353 \\
3 & 423 & 0.3788 & 0.3089 & 0.9639 & 0.3210 \\
4 & 422 & 0.1684 & 0.3028 & 0.8920 & 0.3171 \\
5 & 416 & 0.3050 & 0.3157 & 0.8701 & 0.3163 \\
6 & 394 & 0.4635 & 0.3657 & 0.8089 & 0.3127 \\
7 & 298 & 0.3261 & 0.1849 & 0.7372 & 0.2991 \\
8 & 146 & 0.8587 & 0.1060 & 0.6670 & 0.2996 \\
9 & 42 & 0.8544 & 0.5000 & 0.5849 & 0.2945
\end{tabular}
\end{ruledtabular}
\end{table}

\begin{figure}[h]
\resizebox{0.75\columnwidth}{!}{%
  \includegraphics{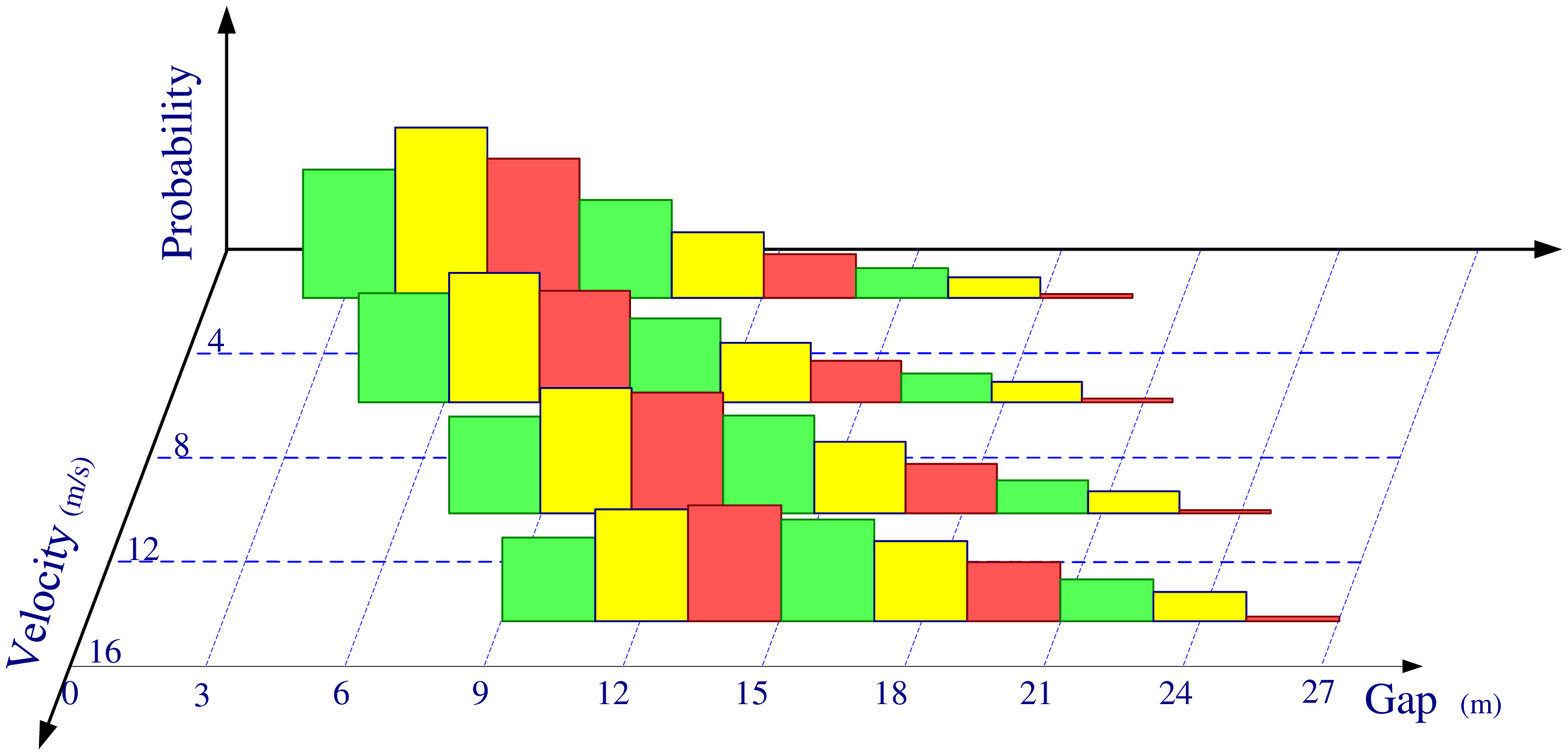}
}
\caption{The assumed discretized steady-state gap distributions
according to different velocities.}
\label{fig:6}       
\end{figure}

By specializing the above Markov-Gap CA model, we can give a
explanation of such phenomena. Here, the length of each cell is set
as $0.1$m. The length of each vehicle is still $4$m. Here, only four
transition matrices are designed; see Fig.~\ref{fig:6}. All the four
distributions are uniformly discretized into 9 states (9 segments,
$2$m per segment) discretized histogram. If the velocity of a
vehicle lies in $[0,4]$m/s, then the gap variation process is
assumed to follows the log-normal distribution with mean $1.9$ and
stand deviation $0.8$. After discretizing, the allowable gap range
is divided into $9$ monospaced states between $2$m to $20$m.
Similarly, we have the other settings as shown in
Table.~\ref{tab:table3}. The maximum coupling distance between two
consecutive vehicles is set as $27$m.

%

\begin{table}[h]
\caption{\label{tab:table3}The designed discretized log-normal type
gap distributions according to different velocities. $\bar{\mu}$ and
$\bar{\sigma}$ denote the mean and standard deviation of the
logarithm of data respectively.}
\begin{ruledtabular}
\begin{tabular}{ccccc}
Index & Velocity range(m/s) & Gap range(m) & $\bar{\mu}$ & $\bar{\sigma}$ \\
\hline
1 & [0,4] & [2,20] & 1.9 & 0.8 \\
2 & [4,8] & [4,22] & 2.2 & 0.7 \\
3 & [8,12] & [7,25] & 2.5 & 0.5 \\
4 & [12,16] & [9,27] & 2.7 & 0.45
\end{tabular}
\end{ruledtabular}
\end{table}

The initial gaps of the vehicles are assigned as random numbers
following a normal distribution ($\mu_{initgap}=1.7$m,
$\sigma_{initgap}=0.1$). The velocity is updated once $T
=0.50\sim0.70$s randomly, because human drivers usually cannot make
decision or take action in a too short time). The maximum velocity
of a vehicle is set to $v_{max}=16$m/s here, and the maximum
ac/decelerating rate is $a_{max}^+=a_{max}^-=6$m/s$^2$. And the
braking parameters are set as $G=0.5$m, $H=12.5$s, $D=6$m/s$^2$.

Before the simulation, all the vehicles are fully stopped. And the
departure of a vehicle starts one by one (a vehicle will start to
move only if the gap ahead is greater than $2.2$m). The first
vehicle will steadily accelerate to $16$m/s in the $16$s
(accelerating rate $1$m/s$^2$) and then keep this speed.

\begin{figure}[h]
\resizebox{0.75\columnwidth}{!}{%
  \includegraphics{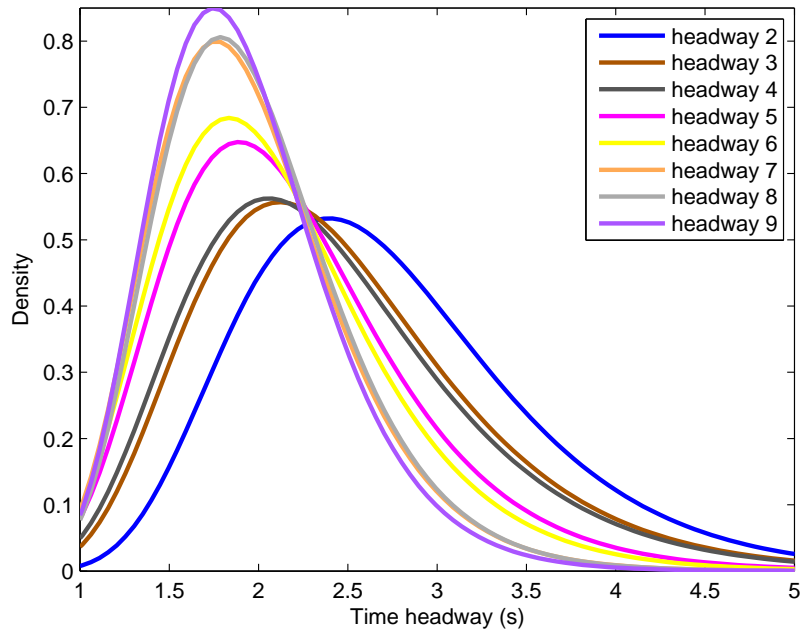}
}
\caption{The distributions of a typical set of departure headways
identified from the histograms obtained in simulation}
\label{fig:7}       
\end{figure}

Results below show that the resulting mean departure headways (400
rounds of simulations for several times) following a similar but
faster decreasing pattern to the empirical ones; see
Fig.~\ref{fig:7}. The statistics of a set of simulated departure
headways are given in Table.~\ref{tab:table4} (the criteria level is
still 95\%). The simulated headways do not decrease exactly like
empirical ones, partly because: 1) the empirical samples for the
last few positions are limited and may be biased, 2) the simple
braking rule (\ref{equ:5}) is still not accurate enough to depict
the real tracking dynamics. More efforts will be put into here for
better concordance between the empirical and simulated data.
However, it is apparent that this model captures the main features
of the driving behaviors in a discharging queue and therefore yields
an accordant macroscopic statistic result through a minimal
description of their microscopic interactions.

%

\begin{table}[h]
\caption{\label{tab:table4}The Kolmogorov-Smirnov (K-S) and
Jarque-Bera (J-B) hypothesis testing results of a typical set of
simulated departure headways£®$\hat{\mu}$ and $\hat{\sigma}$ denote
the mean and standard deviation of the logarithm of data
respectively.}
\begin{ruledtabular}
\begin{tabular}{ccccccc}
Postion & Sample size & K-S test & J-B test & $\hat{\mu}$ & $\hat{\sigma}$ \\
\hline
2 & 400 & 0.7525& 0.0325 & 0.9623& 0.2993 \\
3 & 400 & 0.3678 & 0.0501 & 0.8540 & 0.3211 \\
4 & 400 & 0.6651 & 0.0706 & 0.8283 & 0.3268 \\
5 & 400 & 0.5992 & 0.2564 & 0.7336 & 0.3104 \\
6 & 400 & 0.7980 & 0.3834 & 0.6992 & 0.3035 \\
7 & 400 & 0.8780 & 0.1100 & 0.6413 & 0.2726 \\
8 & 400 & 0.6681 & 0.7080 & 0.6520 & 0.2673 \\
9 & 400 & 0.0912 & 0.1979 & 0.6265 & 0.2595
\end{tabular}
\end{ruledtabular}
\end{table}

Further simulations also reveal that:

1) The general shapes of the simulated departure headways
distributions will be roughly kept according to different
aggregation ratio, unless the number of states is too small. This
proves the Markov Chain Aggregation technique in (\ref{equ:1}) is
applicable.

2) The variation trend of mean departure headways is not very
sensitive to the initial positions and velocities of the vehicles,
which fits the facts observed.

\begin{acknowledgments}

This work was supported in part by National Basic Research Program
of China (973 Project) 2006CB705506, Hi-Tech Research and
Development Program of China (863 Project) 2006AA11Z208, and
National Natural Science Foundation of China 50708055.

\end{acknowledgments}

%

\begin{thebibliography}{}
%
%
\bibitem{ChowdhurySantenSchadschneider2000}
D. Chowdhury, L. Santen, A. Schadschneider, \textit{Phys. Rep.}
\textbf{329}, 199 (2000).

\bibitem{Helbing2001}
D. Helbing, \textit{Rev. Mod. Phys.} \textbf{73}, 1067 (2001).

\bibitem{MahnkeKaupuzsLubashevsky2005}
R. Mahnke, J. Kaupuzs, I. Lubashevsky, \textit{Phys. Rep.}
\textbf{408}, 1 (2005).

\bibitem{Lee2004}
J. W. Lee, \textit{Physica A} \textbf{331}, 531 (2004).

\bibitem{RawalRodgers2005}
S. Rawal, G. J. Rodgers, \textit{Physica A} \textbf{346}, 621
(2005).

\bibitem{KernerKlenovHillerRehborn2006}
B. S. Kerner, S. L. Klenov, A. Hiller, and H. Rehborn, \textit{Phys.
Rev. E} \textbf{73}, 046107 (2006).

\bibitem{Abul-Magd2006}
A.Y. Abul-Magd, \textit{Physica A} \textbf{368}, 536 (2006).

\bibitem{JinSuZhangLi2008}
X. Jin, Y. Su, Y. Zhang, L. Li, arXiv:0803.2619

\bibitem{Kerner2004}
B. S. Kerner, \textit{The Physics of Traffic} (Springer, Heidelberg,
2004).

\bibitem{HelbingTreiberKesting2006}
D. Helbing, M. Treiber, A. Kesting, \textit{Physica A} \textbf{36},
62 (2006).

\bibitem{SchonhofHelbing2007}
M. Schonhof, D. Helbing, \textit{Transp. Sci.} \textbf{41}, 135
(2007).

\bibitem{KrbalekHelbing2004}
M. Krbalek, D. Helbing, \textit{Physica A} \textbf{333}, 370 (2004).

\bibitem{Abul-Magd2007}
A. Y. Abul-Magd, \textit{Phys. Rev. E} \textbf{76}, 057101 (2007).

\bibitem{MaerivoetMoor2005}
S. Maerivoet, B. De Moor, \textit{Phys. Rep.} \textbf{419}, 1
(2005).

\bibitem{NagelScheckenberg1992}
K. Nagel, M. Schreckenberg, \textit{J. Phys. I} \textbf{2}, 2221
(1992).

\bibitem{HelbingSchreckenberg1999}
D. Helbing, M. Schreckenberg, \textit{Phys. Rev. E} \textbf{59},
2505 (1999).

\bibitem{SonKimLee2005}
B. Son, T. Kim, T. Lee, in \textit{Lecture Notes in Computer
Science} \textbf{3481}, 863 (2005).

\bibitem{Luttinen1992}
R. T. Luttinen, \textit{Transp. Res. Rec.} \textbf{1365}, 92 (1992).

\bibitem{MichaelLeemingDwyer2000}
P. G. Michael, F. C. Leeming, W. O. Dwyer, \textit{Transp. Res. F},
\textbf{3}, 55 (2000).

\bibitem{Stewart1994}
W. J. Stewart, \textit{Introduction to the Numerical Solution of
Markov Chains}, (Princeton University Press, Priceton, 1994).

\bibitem{Spears1999}
W. M. Spears, in \textit{Proceedings of the Congress on Evolutionary
Computation}, (1999).

\bibitem{BolchGreinerdeMeerTrivedideMeerTrivedi2006}
G. Bolch, S. Greiner, H. de Meer, K. S. Trivedi, H. de Meer, K. S.
Trivedi, \textit{Queueing Networks and Markov Chains: Modeling and
Performance Evaluation with Computer Science Applications}, 2nd ed.,
(John Wiley and Sons, 2006).

\bibitem{MeynTweedie1994}
S. P. Meyn, R. L. Tweedie, \textit{Markov Chains and Stochastic
Stability}, (Springer-Verlag, New York, 1993).

\bibitem{BoydDiaconisXiao2004}
S. Boyd, P. Diaconis, and L. Xiao, \textit{SIAM Review} \textbf{46},
667 (2004).

\bibitem{Bando1995}
M. Bando, K. Hasebe, A. Nakayama, A. Shibata, Y. Sugiyama,
\textit{Phys. Rev. E}, \textbf{51}, 1035 (1995).

\bibitem{HelbingTilch1998}
D. Helbing, B. Tilch, \textit{Phys. Rev. E} \textbf{58}, 133 (1998).

\bibitem{JiangWuZhu2001}
R. Jiang, Q. S. Wu, Z. J. Zhu, \textit{Phys. Rev. E} \textbf{64},
017101 (2001).

\bibitem{ChangZhangMaoGong2005}
Y. Chang, P. Zhang, L. Mao, Z. Gong, in \textit{Proceedings of
Traffic and Granular Flow 05}, (2005).

\bibitem{LiWang2006}
L. Li, F. Wang, in \textit{Lecture Notes in Computer Science}
\textbf{4153}, 105 (2006).

\bibitem{SuWeiChengYaoZhangLiZhangLi2008}
Y. Su, Z. Wei, S. Cheng, D. Yao, Y. Zhang, L. Li, Z. Zhang, Z. Li,
in \textit{Transportation Research Board Annual Meeting CD}, (2008).

\bibitem{Gardiner1983}
C. W. Gardiner, \textit{Handbook of Stochastic Methods for Physics,
Chemistry and the Natural Sciences}, (Springer-Verlag, New York,
1983).

\end{thebibliography}
%

\end{document}